# Competing ν = 5/2 fractional quantum Hall states in confined geometry

Hailong Fu[a], Pengjie Wang[a], Pujia Shan[a], Lin Xiong[a], Loren N. Pfeiffer[b], Ken West[b], Marc A. Kastner[c,d], and Xi Lin[a,e]

[a]International Center for Quantum Materials, Peking University, Beijing 100871, China

[b]Department of Electrical Engineering, Princeton University, Princeton, New Jersey 08544, USA

[c]Department of Physics, Massachusetts Institute of Technology, Cambridge, Massachusetts 02139, USA

[d]Science Philanthropy Alliance, Palo Alto, California 94306, USA

[e]Collaborative Innovation Center of Quantum Matter, Beijing 100871, China

**Abstract:**

Some theories predict that the filling factor 5/2 fractional quantum Hall state can exhibit non-Abelian statistics, which makes it a candidate for fault-tolerant topological quantum computation. Although the non-Abelian Pfaffian state and its particle-hole conjugate, the anti-Pfaffian state, are the most plausible wave functions for the 5/2 state, there are a number of alternatives with either Abelian or non-Abelian statistics. Recent experiments suggest that the tunneling exponents are more consistent with an Abelian state rather than a non-Abelian state. Here, we present edge-current-tunneling experiments in geometrically confined quantum point contacts, which indicate that Abelian and non-Abelian states compete at filling factor 5/2. Our results are consistent with a transition from an Abelian state to a non-Abelian state in a single quantum point contact when the confinement is tuned. Our observation suggests that there is an intrinsic non-Abelian 5/2 ground state, but that the appropriate confinement is necessary to maintain it. This observation is important not only for understanding the physics of the 5/2 state, but also for the design of future topological quantum computation devices.

**Significance Statement:**

The 5/2 fractional quantum Hall state has captured the fascination of the scientific and non-scientific community, because of proposals to use it for quantum computation. These proposals have been based on the premise that the quasi-particles in this state have non-Abelian statistics. Experiments supporting both Abelian 5/2 state and non-Abelian 5/2 state have been reported, and no consensus has been reached. Using measurements of tunneling between edge states, we suggest that both the Abelian and non-Abelian states can be stable in the same device, but under different conditions. Our discovery resolves the inconsistencies between existing experiments. More importantly, the knowledge of the competition between ground states is important for maintaining non-Abelian statistics in application to topological quantum computation.

**Introduction:**

Among the roughly 100 known fractional quantum Hall (FQH) states, the filling factor $\nu = 5/2$ state is special. It is an even-denominator state with elementary excitations that may have non-Abelian statistics (1-10). If the ground state is non-Abelian, it would be insensitive to environmental decoherence (11) and would therefore be useful for fault-tolerant topological quantum computation. Because of this potential application, much theoretical effort has focused on the 5/2 state, and a variety of wave functions have been proposed (1-3, 5-7, 12-15). There have also been several experiments on the 5/2 state, with more of them supporting non-Abelian than Abelian statistics (16, 17). All of the proposed wave functions for the 5/2 state have an effective charge $e*$ of the quasiparticles that is a quarter of the elementary charge e, and this effective charge has been confirmed experimentally (18-21). However, the wave function of the 5/2 state is still under discussion.

The proposed wave functions of the 5/2 state can be distinguished by the strength of the interaction between quasiparticles, described by a coupling constant $g$. This coupling constant can be obtained from experiments using weak-tunneling theory (22).

The FQH effect emerges in a highly interacting 2D electron gas (2DEG) at high magnetic field and ultralow temperature. The FQH states have gapless conducting edge currents at the boundaries of the 2DEG, in which fractionally charged quasiparticles carry the current. If the counter-propagating edge currents of a FQH state are brought close enough together in a quantum point contact (QPC), back-scattering is induced. In the weak-tunneling regime, the rate of quasiparticle tunneling depends on the small voltage difference between the edges, and scales with temperature as a function of $e*$ and $g$. The proposed wave functions for 5/2 state have different values of $g$, so the determination of $g$ can discriminate among them. The existing tunneling experiments have demonstrated the validity of weak-tunneling theory in the $\nu = 5/2$ FQH state, but one is consistent with the non-Abelian anti-Pfaffian wave function (23), whereas others indicate the Abelian 331 wave function (24, 25).

We have measured the tunneling conductance between edges of the 5/2 state in a QPC with adjustable confinement conditions, and we have found signatures of both non-Abelian and Abelian wave functions. The tunneling conductance has been measured as a function of the edge voltage difference and temperature, and by fitting the tunneling conductance to weak-tunneling theory, the strength of the interaction $g$ and the effective charge $e*$ of the 5/2 state have been extracted. We find that $g$ depends on the voltage applied to the gates of the QPC.

**Results:**

The separation of the two fingers forming the QPC is 600 nm for the results presented in the main text, but data from additional QPC geometries can be found in Figs. S1 and S2. The width of the QPC finger is 200 nm, similar to that used in the previous works (23, 25). The 2DEG is 210 nm below the QPC and the GaAs/AlGaAs surface, with a density of $2.9\times10^{11}$ cm$^{-2}$ and a mobility of $2.3\times10^{7}$ cm$^2$ V$^{-1}$ s$^{-1}$. The energy gap of the 5/2 state is 148 mK calculated from the temperature dependence of the longitudinal resistance $R_{XX}$, which is thermally activated.

A quantized Hall resistance $R_{XY}$ plateau at filling factor 5/2 is shown in Fig. 1,

with a near-zero $R_{XX}$, indicating a well-established FQH state. When a negative enough voltage is applied to the gates of the QPC, the edge currents are brought close enough together to induce tunneling. As a result, the Hall resistance across the QPC, the diagonal resistance $R_D$, increases above the quantized value. The tunneling conductance is calculated as $g_T = \dfrac{R_D - R_{XY}}{R_{XY}^2}$, and its dependence on the bias current $I_{DC}$ and the temperature has been predicted by weak-tunneling theory (22, 23). To compare the experiment with theory, one must use the temperature of the electrons, rather than the lattice temperature, which is typically measured using a resistor thermometer and called the refrigerator temperature. The temperatures given in this work is the electron temperature, unless otherwise specified, and their measurement is discussed in *Methods*.

There is strong evidence that the electron density inside and outside the QPC is very similar in our experiment. $R_{XY}$ and $R_D$ as a function of magnetic field are plotted in Fig. 1; that features for the integer quantum Hall effect (IQHE) and the re-entrant IQHE occur at the same fields guarantees uniform density, and therefore constant filling factor, throughout the entire device. This behavior is found in all QPC voltages ($V_{QPC}$) studied in this work. This uniformity is realized through gate annealing (23) (see *Methods* for experimental details), and it greatly simplifies the theoretical analysis of the edge physics. The uniform density also implies that the confinement of the 2DEG at its boundaries is very sharp. The 2DEG density within the QPC is usually significantly smaller than the bulk 2DEG density, unless the device is annealed at ~4 K with gate voltage applied (23).

At $V_{QPC}$ = -1.50 V (Fig. 2*A*) a single pair of $e^*/e = 0.25$ and $g = 0.52$ fit the data well over a wide temperature range, which indicates a stable and unvarying 5/2 state wave function under an appropriate confinement condition. Here, the diagonal resistance $R_D$ as a function of bias current $I_{DC}$ is fitted at different temperatures simultaneously to extract $e^*$ and $g$, which is same as the earlier tunneling works (23-25). The fitting equation originating from weak-tunneling theory and the fitting details can be found in *Methods*. As mentioned above, the effective charge $e^*$ has been predicted to be e/4 in all proposed 5/2 wave functions, and our measurements are consistent with this value. On the other hand, different wave functions have different $g$ values, and we compare those values with our results in Fig. 2$B$. The theoretical predictions of pairs of $e^*$/e and $g$ for six theoretical candidate states are mapped, with red circles for non-Abelian candidates and black squares for Abelian states. Inside the contour "1", where the average fitting-error residual is smaller than the experimental noise, we find only the point $(e^*$/e, $g) = (1/4, 1/2)$, which corresponds to either the non-Abelian anti-Pfaffian (5, 6) or the non-Abelian $U(1) \times SU_2(2)$ state (1, 2). The Pfaffian and anti-Pfaffian states represent a chiral p-wave topological superconductor of composite fermions, and the vortices of which support Majorana bound states obeying non-Abelian statistics (5, 6). The non-Abelian $U(1) \times SU_2(2)$ state (2) was motivated by the parton model (1), in which electrons are broken into fictitious particles called partons. We find similar evidence for states corresponding to (1/4, 1/2) in two other QPCs with widths of 400 and 800 nm (Figs. S1 and S2). The gate voltage used for the 400 nm is -1.2 V in Fig. S1 and that for the 800 nm is -2.3 V in Fig. S2. A detailed analysis of how $e^*$ and $g$ are extracted from such data can be found in *Methods*.

**Discussion:**

We find that the value of $g$ changes when the confinement condition is tuned. At a more negative $V_{QPC}$, $g$ decreases significantly below 1/2, and interestingly, there is a temperature dependence of $g$. As shown in Fig. 3$A$, if we fit the lowest temperature and the highest temperature of tunneling data, separately, with $e^*$/e fixed at 1/4, the values of $g$ for $V_{QPC} = -1.65$ V are different. This variation of g is to be contrasted with $V_{QPC} = -1.50$ V, at which there is almost no temperature dependence of $g$. Note that at the lowest temperature, the line shapes are strikingly different at the two gate voltages, with an undershoot present only at $V_{QPC} = -1.65$ V, which is nicely reproduced by the theory only with the smaller value of $g$. We plot the $g$ values at $V_{QPC} = -1.50$ V and $V_{QPC} = -1.65$ V as a function of temperature in Fig. 3$B$, illustrating qualitative as well as quantitative differences when the confinement is changed. For completeness, we have carried out the analysis for $V_{QPC} = $ -1.65 V over the same temperature range (Fig. S3) as in Fig. 2, which leads to $e^*/e = 0.17$ and $g = 0.30$, similar to the previously reported results, consistent with the Abelian 331 state (24, 25), which was the Halperin's generalization of Laughlin's wave function for a bilayer system. More data for the full temperature range fits and single temperature fits at individual values $V_{QPC}$ can be found in Figs. S4 and S5.

To further investigate the confinement's influence on the 5/2 state, we have analyzed the tunneling data at each temperature and value of $V_{QPC}$ (Fig. 4). One sees the region of $g \sim 0.5$, the value associated with the non-Abelian state. Under stronger confinement, the $g$ value (darker colored region labeled as "Abelian") is close to, but slightly larger than the number 3/8 predicted for the 331 state. There is stronger support for the 331 state in other tunneling experiments (24, 25) than our data. The deviation from 3/8 has been proposed theoretically (26) to result from influence of device geometry.

Our results (Figs. 3 and 4) strongly suggest that the non-Abelian state is favored at weak confinement, whereas the Abelian state is more stable at strong confinement. We infer from this observation that in the bulk of 2DEG without any confinement, the intrinsic state for 5/2 is non-Abelian. The suppression of the non-Abelian state by strong confinement has been predicted by theoretical calculations (27), but the underlying mechanism for the suppression of the non-Abelian state is not clear. We cannot completely exclude a very small variation of electron density, smaller than our experimental resolution, and this density variation might affect which state is the most stable. There has been a nonmonotonic energy gap as a function of density in an open geometry sample, which is explained as competing 5/2 states, one spin-polarized and one spin-unpolarized when the density is tuned: it is argued that the electron interaction is changed through Landau level mixing (28). The related studies at the 5/2 state have found much stronger evidence for spin-polarization than spin-unpolarization in open geometry samples (29-37). It should be noted that such an energy gap measurement (28) was studied at magnetic field lower than 1 T, and another measurement at magnetic field similar to this work only showed monotonic behavior between the energy gap and the density (38). Lastly, the electrostatic potential caused by the gates will certainly not be constant over the width of the confined region, and its shape may influence the relative stability of the two states.

In summary, our results support that the non-Abelian state is the proper description of the 5/2 state under weak confinement. More importantly, the indication of that two different states compete in confined geometries enriches the understanding of the 5/2 FQH state, as well as resolves some apparent inconsistencies between existing measurements using various techniques. Knowing how to stabilize the non-Abelian state will be important for interferometer experiments to directly test the quasiparticle statistics and in applications to topological quantum computation.

**Methods:**

1. Fabrication and transport measurement

A Hall bar with a width of 150 $\mu$m was shaped by wet etching of 240:8:1 $H_2O:H_2O_2:H_2SO_4$ solution for 120 s after coating with photoresist AZ1500 and photolithography. Ohmic contacts were made of Pt/Au/Ge alloy deposited by e-beam evaporation and annealed at 550℃ for 100 s. QPCs were formed by two Cr/Au top gates deposited by e-beam evaporation after lithography. Applying a negative voltage to the gates depleted the electrons underneath at $V_{QPC} \sim$-1.3 V, and more negative voltage brought the opposite edges closer, before the QPC was completely pinched-off. QPCs were annealed at -1.6, -1.8, and -2.5 V (400, 600 and 800nm, respectively) near 4 K.

The transport experiments were carried out in a dilution refrigerator (MNK126-450; Leiden Cryogenics BV) with a base refrigerator temperature below 6 mK. The main thermometry was calibrated against a noise thermometer and checked by a cerium magnesium nitrate susceptibility thermometer. The lowest electron temperature for this study was 18 mK determined by measuring the temperature dependence of the longitudinal resistance $R_{XX}$ at an appropriate FQH state. Above 20 mK, the electron temperature was equal to the refrigerator temperature confirmed by the temperature dependence of FQH states' $R_{XX}$. The measurements were performed by using standard lock-in techniques, with an AC excitation of 1 nA at 6.47 Hz.

2. Data analysis

The effective charge $e^*$ and the strength of the interaction $g$ can be extracted from the following equations (22, 23):

$$g_T(T, I_{DC}) = AT^{(2g-2)}F(g, \frac{e^* I_{DC} R_{XY}}{k_B T})$$

$$F(g, x) = B(g + i\frac{x}{2\pi}, g - i\frac{x}{2\pi})\left\{\pi \cosh(\frac{x}{2}) - 2\sinh(\frac{x}{2})\,\text{Im}\left[\Psi(g + i\frac{x}{2\pi})\right]\right\}$$

where $g_T = \dfrac{R_D - R_{XY}}{R_{XY}^2}$, $B(x, y)$ represents the Euler beta function, and $\Psi(x)$ is the digamma function. Because there is only a scale factor and an offset difference between $R_D$ and $g_T$, we directly fitted $R_D$ to obtain $e^*$ and $g$. The line shape of the tunneling conductance as a function of bias current is sufficiently complex in this equation that $g$ and $e^*$ are well determined at any temperature and gate voltage if weak-tunneling theory applies. By fixing $e^*$ and $g$, we fit $R_D$ versus $I_{DC}$ at different temperatures simultaneously to verify the agreement between the 5/2 state's theories and our data in Fig. 2B. Except where noted the three parameters in this equation, $A$, $e^*$ and $g$ are allowed to vary to get the best least-squares fit. Because $g_T$ is calculated from the diagonal resistance and there is a small hysteresis in $I_{DC}$, two more free parameters are added to offset the resistance and the bias current, but these are substantially independent of temperature. The reliability of the fitting is checked by the normalized fitting error, defined as the fit residual divided by the experimental noise $2 \times 10^{-4}$ h/e$^2$.

$g$ can be calculated from a proposed wave function and is a discrete number. In an ideal experiment, we would find one of the $g$ values predicted by theory. In real measurements, there may be deviations from the theoretical values resulting from the influence of device geometry (26). Nonetheless, because the total number of known 5/2 wave functions is limited, we can exclude some candidate wave functions. In this work, $g$ is very close to 1/2 at $V_{QPC}$ = -1.50 V and it is very close to 3/8 at $V_{QPC}$ = -1.65 V. Among the theoretically predicted states, there are only two with $g = 1/2$, which are non-Abelian, and only one with $g = 3/8$, which is Abelian.

**Acknowledgements:**

We thank Rui-Rui Du, D. E. Feldman, Jainendra K. Jain, Woowon Kang, Dung-hai Lee, Xin Wan, Fa Wang, Xiao-Gang Wen, Xin-Cheng Xie and Fuchun Zhang for discussions. The work at PKU was funded by NSFC (Grant No. 11274020 and 11322435) and NBRPC (Grant No. 2015CB921101 and 2012CB921301). The work at Princeton University was funded by the Gordon and Betty Moore Foundation through the EPiQS initiative Grant GBMF4420, by the National Science Foundation MRSEC Grant DMR-1420541, and by the Keck Foundation.

**Author Contributions**:

H.F., P.W., P.S. and X.L. performed the measurements. H.F. fabricated the devices. L.N.P. and K.W.W. prepared and supplied the GaAs wafer. M.A.K. and X.L. initiated the research. H.F., P.W., L.X. and X.L. analyzed the results. H.F., M.A.K. and X.L. discussed the results and wrote the manuscript.

**Conflict of interest:**

The authors declare no competing financial interests.

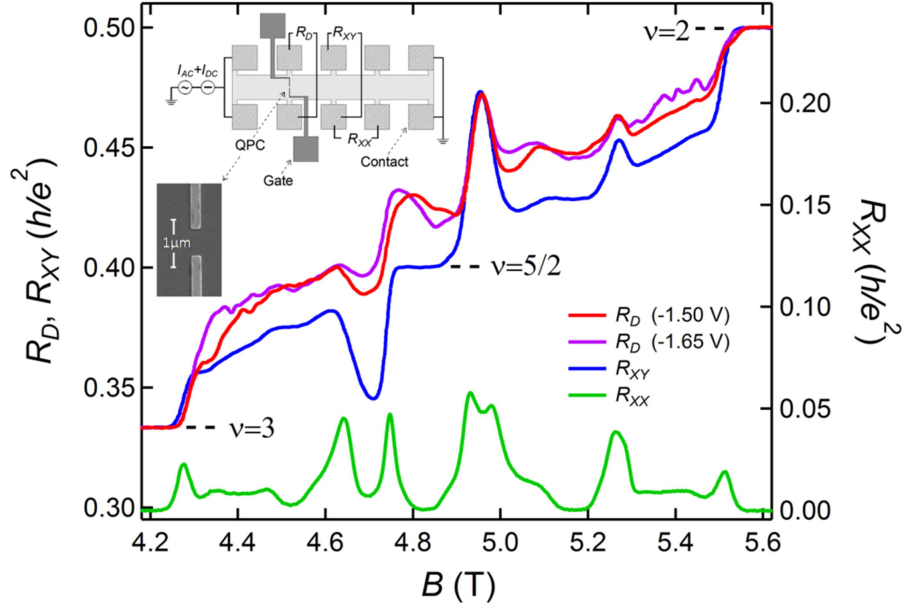

Fig. 1. Magnetic field dependence of the longitudinal resistance $R_{XX}$, the Hall resistance $R_{XY}$, and the diagonal resistance $R_D$ in the second landau level. When the filling factor is equal to 5/2, $R_{XY} = \dfrac{2h}{5e^2}$, which is marked by a dash line. The dip and peak of $R_{XY}$ on both sides of 5/2 are reentrant integer quantum Hall features, signifying a well-developed 5/2 state. Due to the tunneling, $R_D$ is larger than $R_{XY}$ at $\nu = 5/2$. Resistances were measured at a refrigerator temperature of 6 mK, the lowest temperature of our equipment and corresponding to an electron temperature of ~18 mK. The inset is a sketch of the Hall bar device and the measurement setup. $I_{DC}$ bias generates the voltage difference between counter-propagating edges. A scanning electron micrograph of a different device with identical designed gate geometry is also provided.

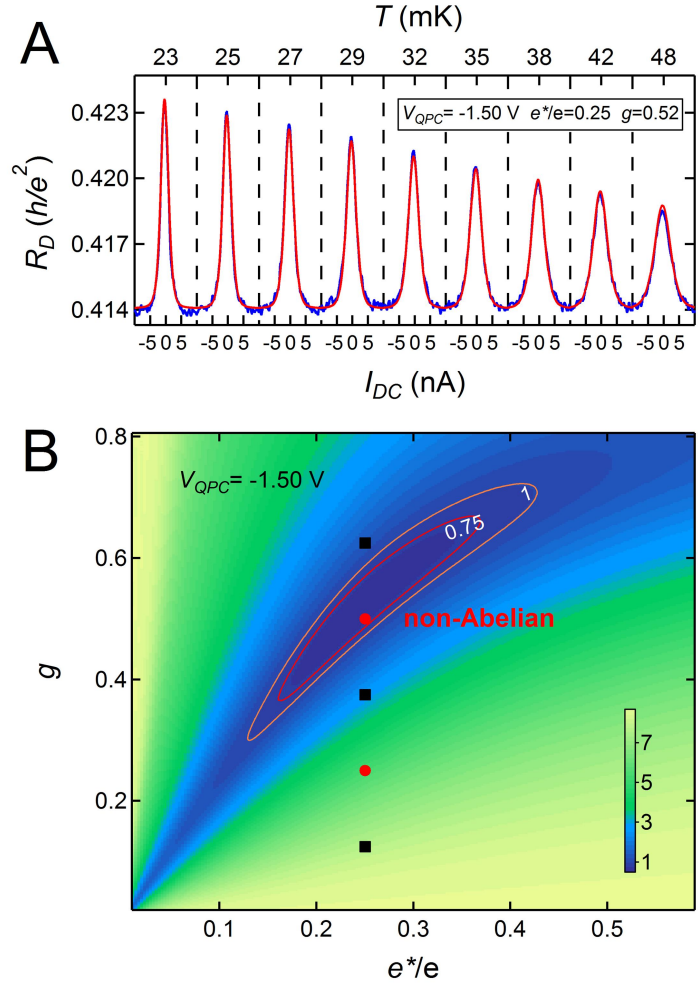

Fig. 2. Tunneling results at the ν = 5/2 state, $V_{QPC}$ = -1.50 V. (*A*) Direct current (DC) bias and temperature dependence of $R_D$ (blue line) and the least-squares fitting curve (red line) at magnetic field B = 4.80 T. $R_D$-$I_{DC}$ curves at different temperatures are fitted simultaneously. From the best fit, $e*/e$ = 0.25 and $g$ = 0.52. (*B*) Fit residual divided by the experimental noise as a function of fixed pair of ($e*/e$, $g$). The darker color represents better agreement between the experimental data and the theoretical prediction. Pairs of ($e*/e$, $g$) of some proposed non-Abelian states (red circles) and Abelian states (black squares) are marked [$g$ = 1/8, K = 8 state (12); $g$ = 1/4, Pfaffian state (3); $g$ = 3/8, 331 state (13, 14); $g$ = 1/2, anti-Pfaffian state (5, 6) or U(1) × SU$_2$(2) state (1, 2); $g$ = 5/8, anti-331 state (13, 14, 26)]. The contour numbers (1 and 0.75) are defined as the average fitting error residual divided by the experimental noise.

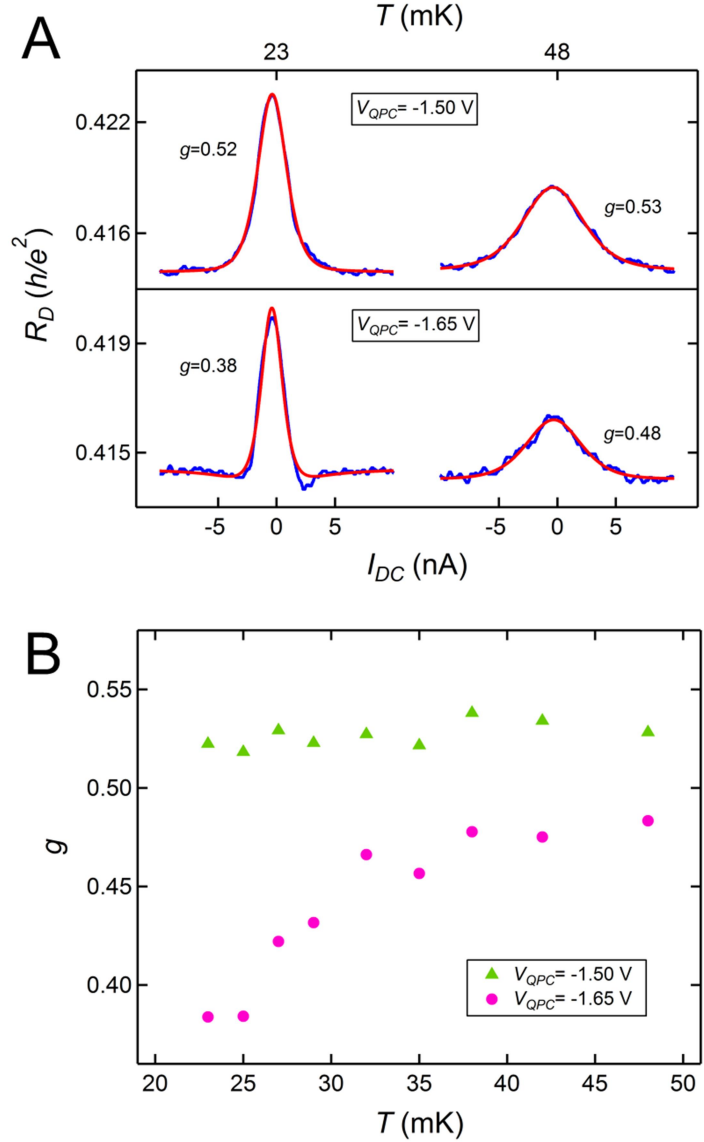

Fig. 3. Temperature dependence of *g* at $V_{QPC}$ = -1.50 V and $V_{QPC}$ = -1.65 V with *e\**
fixed at e/4. (*A*) DC bias dependence of $R_D$ (blue lines) and the least-squares fitting
curves (red lines) at B = 4.80 T. $R_D$-$I_{DC}$ curves at 23 and 48 mK are fitted individually.
(*B*) Values of *g* from individual temperature fits at $V_{QPC}$ = -1.50 V and $V_{QPC}$ = -1.65 V.

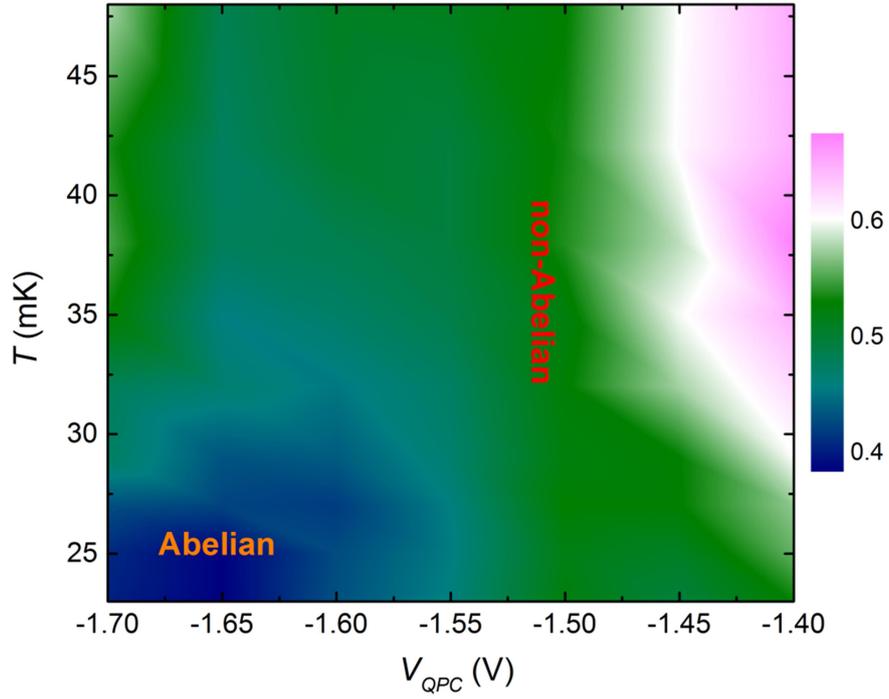

Fig. 4. $g$ of the $\nu = 5/2$ state in a QPC as a function of temperature and confinement from the tunneling measurements. The color represents the value of coupling constant $g$, which is extracted from $R_D$-$I_{DC}$ curves with $e*$ fixed at e/4. The $g$ value matrix from the temperature (23, 25, 27, 29, 32, 35, 38, 42, 48 mK) and the $V_{QPC}$ (-1.70, -1.65, -1.60, -1.55, -1.50, -1.45, -1.40 V) discrete measurements is smoothed for clarity.

# Supporting information

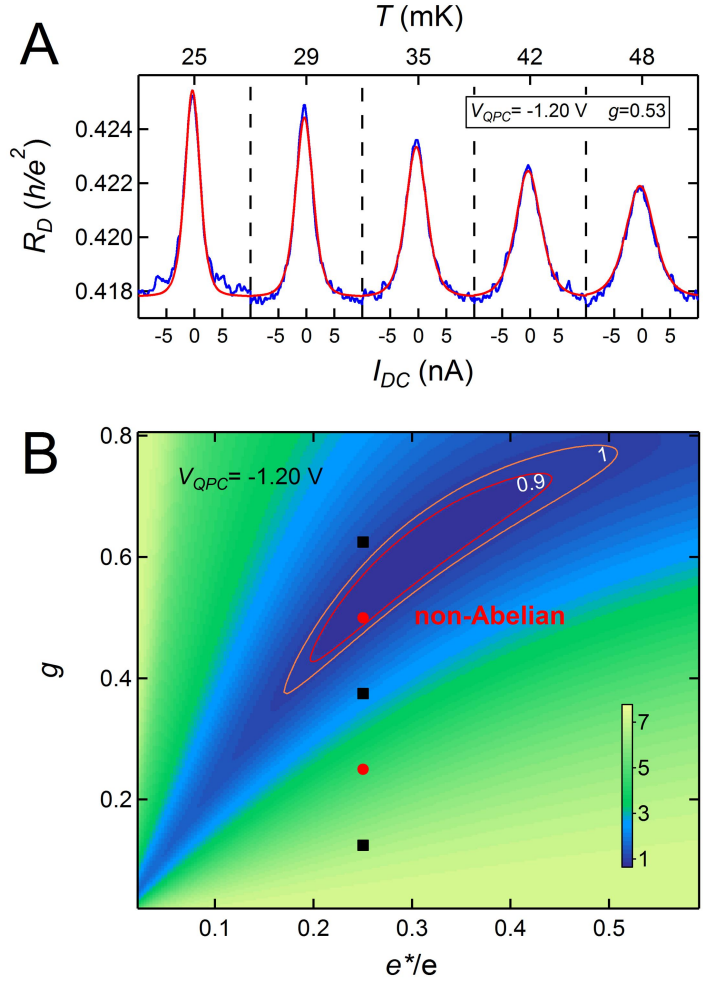

Fig. S1. Tunneling results at the ν = 5/2 state of the 400-nm QPC device. (*A*) DC bias and temperature dependence of $R_D$ (blue line) and the least-squares fitting curve (red line) at magnetic field *B* = 4.80 T. $R_D$-$I_{DC}$ curves at different temperatures are fitted simultaneously. At $V_{QPC}$ = -1.20 V, *g* in the best fit is 0.53 with *e** fixed at e/4. The free parameter fitting generates *e**/e = 0.29 and *g* = 0.59. (*B*) Fit residual divided by the experimental noise as a function of fixed pair of (*e**/e, *g*) at $V_{QPC}$ = -1.20 V. Pairs of (*e**/e, *g*) of some proposed non-Abelian states (red circles) and Abelian states (black squares) are marked [*g* = 1/8, K = 8 state (12); *g* = 1/4, Pfaffian state (3); *g* = 3/8, 331 state (13, 14); *g* = 1/2, anti-Pfaffian state (5, 6) or U(1) × SU₂(2) state (1, 2); *g* = 5/8, anti-331 state (13, 14, 26)]. The contour numbers (1 and 0.9) are defined as the average fitting error residual divided by the experimental noise.

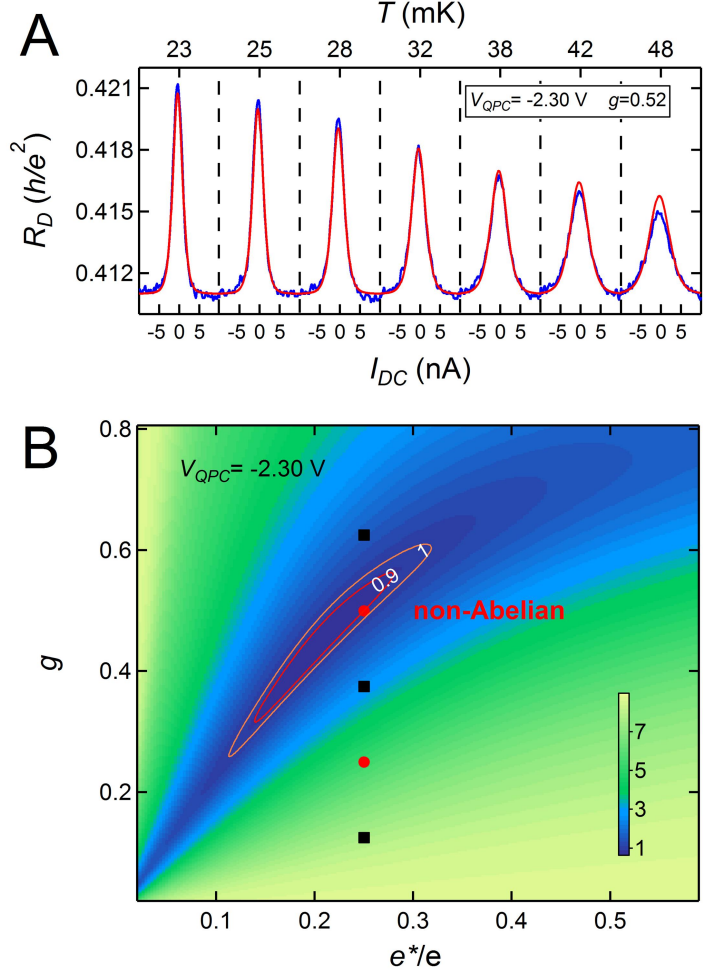

Fig. S2. Tunneling results at the ν = 5/2 state of the 800-nm QPC device. (*A*) DC bias and temperature dependence of $R_D$ (blue line) and the least-squares fitting curve (red line) at magnetic field $B$ = 4.80 T. $R_D$-$I_{DC}$ curves at different temperatures are fitted simultaneously. At $V_{QPC}$ = -2.30 V, $g$ in the best fit is 0.52 with $e^*$ fixed at e/4. The free parameter fitting generates $e^*$/e = 0.19 and $g$ = 0.42. (*B*) Fit residual divided by the experimental noise as a function of fixed pair of ($e^*$/e, $g$) at $V_{QPC}$ = -2.30 V. Pairs of ($e^*$/e, $g$) of some proposed non-Abelian states (red circles) and Abelian states (black squares) are marked [$g$ = 1/8, K = 8 state (12); $g$ = 1/4, Pfaffian state (3); $g$ = 3/8, 331 state (13, 14); $g$ = 1/2, anti-Pfaffian state (5, 6) or U(1) × SU$_2$(2) state (1, 2); $g$ = 5/8, anti-331 state (13, 14, 26)]. The contour numbers (1 and 0.9) are defined as the average fitting error residual divided by the experimental noise.

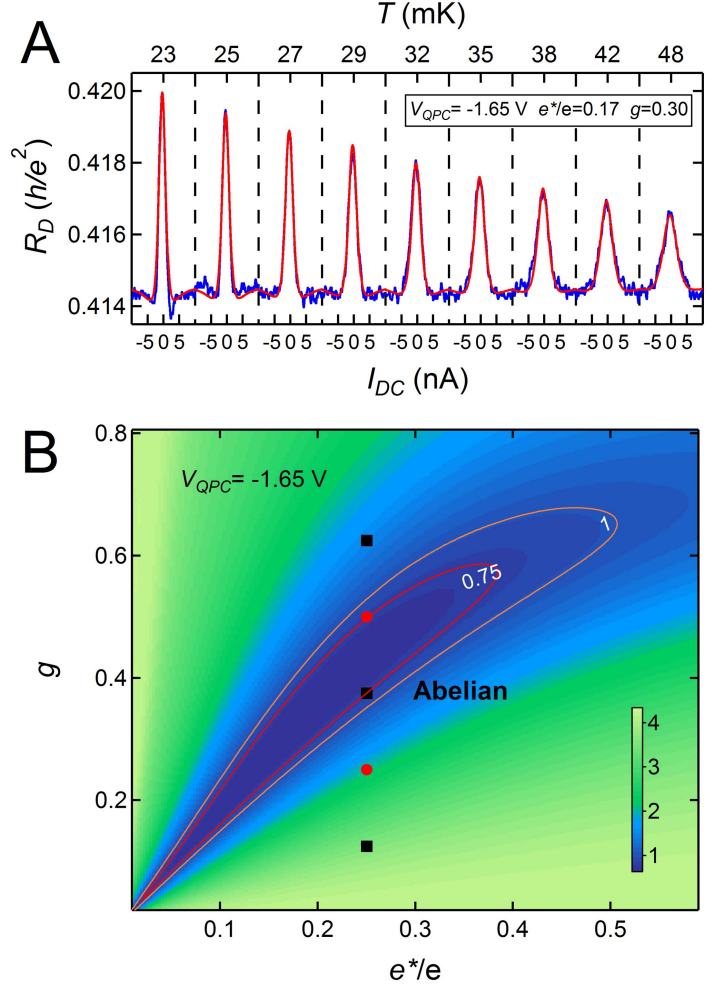

Fig. S3. Tunneling results at the v = 5/2 state of the 600-nm QPC device, $V_{QPC}$ = -1.65 V. (*A*) DC bias and temperature dependence of $R_D$ (blue line) and the least-squares fitting curve (red line) at magnetic field $B$ = 4.80 T. $R_D$-$I_{DC}$ curves at different temperatures are fitted simultaneously, although at this gate voltage, as discussed in the *Discussion*, there is evidence that $g$ depends on temperature. From the best fit, $e*/e$ = 0.17 and $g$ = 0.30. (*B*) Fit residual divided by the experimental noise as a function of fixed pair of ($e*/e$, $g$). The darker color represents the better agreement between the experimental data and the theoretical prediction. Pairs of ($e*/e$, $g$) of some proposed non-Abelian states (red circles) and Abelian states (black squares) are marked [$g$ = 1/8, K = 8 state (12); $g$ = 1/4, Pfaffian state(3); $g$ = 3/8, 331 state (13, 14); $g$ = 1/2, anti-Pfaffian state (5, 6) or U(1) × SU$_2$(1, 2) state (1); $g$ = 5/8, anti-331 state (13, 14, 26)]. The contour numbers (1 and 0.75) are defined as the average fitting-error residual divided by the experimental noise.

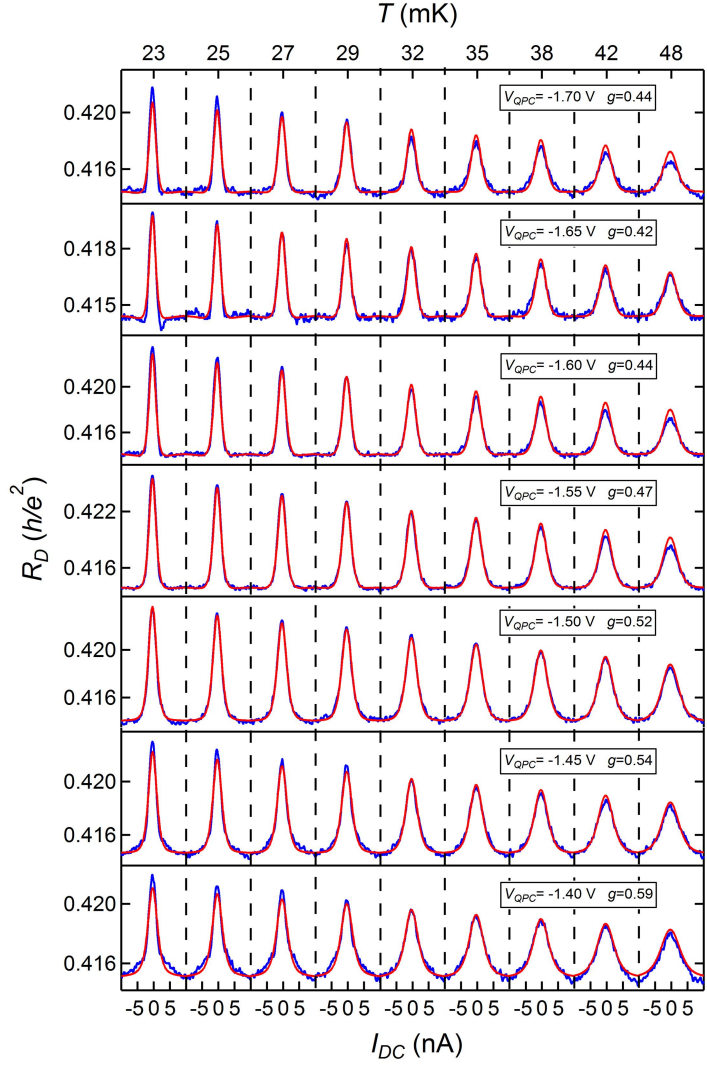

Fig. S4. The least-squares fitting results for the 600-nm QPC device with $e^*$ fixed at e/4. From up to down, the gate voltages ($V_{QPC}$) are: -1.70, -1.65, -1.60, -1.55, -1.50, -1.45 and -1.40 V, respectively. The full temperature fits with $V_{QPC}$= -1.50 V and $V_{QPC}$ = -1.65 V are better than the others.

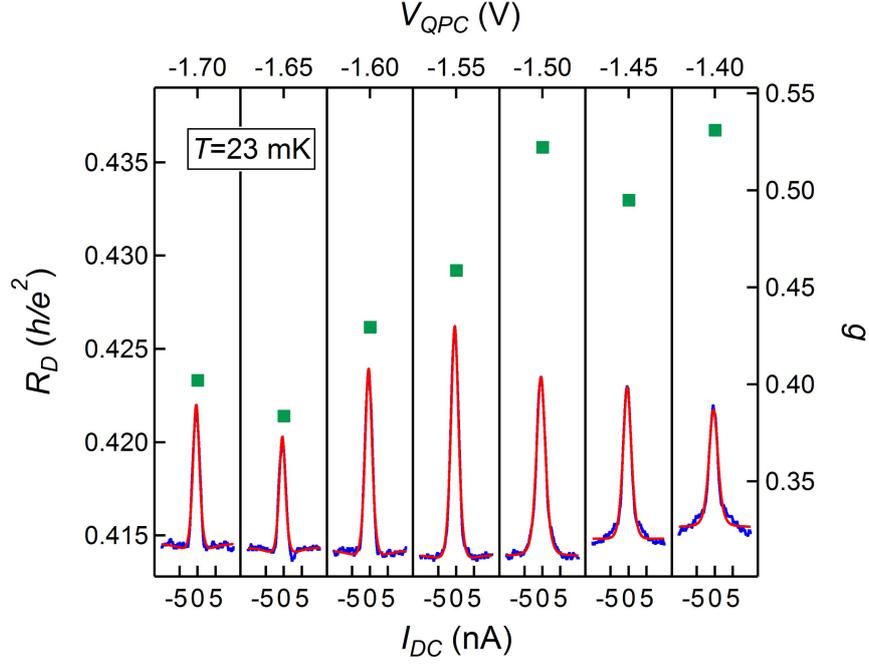

Fig. S5. DC bias dependence of $R_D$ at 23 mK and gate dependence of the fitted $g$ values from the 600-nm QPC device. Blue lines are $R_D$ measured at 23 mK as a function of bias current $I_{DC}$. $R_D$-$I_{DC}$ curves for different voltages are plotted together for comparison. The red lines are individual fits based on weak-tunneling theory with fixed $e^* = e/4$. The fitted $g$ values (green squares and right axis) are plotted as a function $V_{QPC}$. Note the qualitative difference in the peak shapes, with a distinct dip on the side of the main peak for $g = 0.38$ ($V_{QPC} = $ -1.65 V) and no such dip at $g = 0.52$ ($V_{QPC} = $ -1.50 V).